\documentclass[12pt]{revtex4}
\usepackage{bm}

\begin{document}
\preprint{APS/123-QED}
\title{Scissors modes in triaxial metal clusters}
\author{P.-G. Reinhard$^1$, V.O. Nesterenko$^{1,2}$, 
E. Suraud$^3$, S. El Gammal$^4$, and W. Kleinig$^{2,5}$}
\affiliation{
$^1$ Institut fur Theoretische Physik,Universitat Erlangen,
D-91058 Erlangen, Germany,
E-mail: mpt218@theorie2.physik.uni-erlangen.de
\\
$^2$ Bogoliubov Laboratory of Theoretical Physics,
Joint Institute for Nuclear Research
141980, Dubna, Moscow Region, Russia,
E-mail: nester@thsun1.jinr.ru\\
$^3$ Laboratoire de Physique Quantique,
 Universit{\'e} Paul Sabatier,
 118 Route de Narbonne, 31062 Toulouse, cedex,
\\
$^4$ Physics Department, Faculty of Science, El-Menoufia 
University, Shebin El-Kom, Egypt
\\
$^5$
Technische Universit$\ddot a$t Dresden,
Institut f$\ddot u$r Analysis, Dresden, D-01062, Germany
}
\date{\today}

\begin{abstract}
We study the scissors mode (orbital M1 excitations) in small 
Na clusters, triaxial metal clusters ${\rm Na}_{12}$ and 
${\rm Na}_{16}$ and the close-to-spherical ${{\rm Na}_9}^+$, 
all described in DFT with detailed ionic background.
The scissors modes built on spin-saturated ground and 
spin-polarized isomeric states are analyzed in virtue of both 
macroscopic collective and microscopic shell-model treatments. 
It is shown that the mutual destruction of Coulomb and the 
exchange-correlation parts of the residual interaction makes 
the collective shift small and the net effect can depend on 
details of the actual excited state.  The crosstalk with 
dipole and spin-dipole modes is studied in detail. In 
particular, a strong crosstalk with spin-dipole
negative-parity mode is found in the case of spin-polarized 
states. Triaxiality and ionic structure considerably complicate 
the scissors response, mainly at expense of stronger 
fragmentation of the strength. Nevertheless, even in these 
complicated cases the scissors mode is mainly determined by the 
global deformation. The detailed ionic structure destroys the 
spherical symmetry and can cause finite M1 response (transverse 
optical mode) even in clusters with zero global deformation. But 
its strength turns out to be much smaller than for the genuine 
scissors modes in deformed systems.
\end{abstract}

\maketitle

\section{Introduction}

The scissors mode in nuclei is a small-angle collective rotation 
of a {\it deformed} proton cloud against the complementing 
deformed neutron cloud. 
It thus belongs to the basic magnetic orbital dipole excitations. 
A rotation-like character of the mode assumes that it exists only 
in the systems with broken spherical symmetry.    
Early theoretical predictions are found in Refs. \cite{Iu_M1,LS_M1} 
and its first experimental observation was reported in Ref. 
\cite{richter}. For a review see Ref. \cite{I97}. Since then, the 
scissors mode has also been discussed in
other deformed systems, for example in quantum dots \cite{BE_QD},
Bose-Einstein condensates \cite{BE_M1}, and last not least in metal
clusters \cite{lipparini_PRL,Ne_PRL}. In the case of deformed metal 
clusters, the mode is viewed as a 
rotation-like collective oscillations of the electron cloud against 
the ionic background. There exist already a few theoretical 
considerations on that topic, early estimates from a collective 
perspective \cite{lipparini_PRL}, RPA calculations 
with a phenomenological Woods-Saxon potential \cite{Ne_PRL},  
and fully detailed RPA calculations using density-functional theory 
and jellium for the ionic background \cite{Ne_next}. The relation of 
scissors modes to orbital magnetism had been discussed in Refs. 
\cite{lipparini_PRL,Ne_PRL,Ne_next,Ne_Rich,FPR}. Experimental access 
to scissors modes in clusters is yet an unresolved problem 
\cite{Duval} (see also discussion in Ref. \cite{Ne_next}). 
Nevertheless, 
or just because of that, further theoretical investigations are 
worthwhile in order to elucidate the nature of these modes and 
their key features. 

It is the aim of this paper to present a theoretical investigation of
scissors mode in Na clusters using time-dependent density functional
theory for the electrons and fully detailed ionic background. We will
consider in particular the triaxial clusters ${{\rm Na}_{12}}$ and
${{\rm Na}_{16}}$. These two test cases provide several interesting
features.  As triaxial clusters, they have deformation in all three
spatial directions and thus should display three distinct scissors
modes. On the other hand, the single-particle excitation spectra are
most diffuse in these soft systems which may give rise to strong
fragmentation. Both clusters have spin-polarized and axially symmetric
isomers which give rise to crosstalk with the spin-dipole mode which
is energetically very close to the scissors mode.  (The dipole and
spin-dipole excitation spectra in ${{\rm Na}_{12}}$ had been studied
extensively in Ref. \cite{spin_na12}.) The axial isomers allow also a
direct comparison between scissors modes in axial versus triaxial
shapes. We will investigate the impact of ionic structure
on the scissors modes in two ways: first, we compare the
triaxially deformed test cases with the results of the smooth
jellium model, and second, we consider as a complementing example the
nearly spherical ${{\rm Na}_9}^+$ with full ionic structure. 
The test case of axially symmetric ${{\rm Na}_{11}}^+$ with jellium
background is used to disentangle the various (counteracting)
parts of the interaction, which compose the final prediction of the
frequency of the mode. 

\section{Framework}
\label{sec:frame}

The electron cloud of the clusters is described by density-functional
theory (DFT) at the level of the time-dependent local-density approximation
(TDLDA) using actually the density functional of Ref. \cite{GunLun}. The
coupling of ions to electrons is described by a local pseudo-potential
which has been proven to yield correct ground state and optical
excitation properties of simple sodium clusters \cite{kuemmel}.
The electronic wavefunctions are represented on an equidistant grid in
three-dimensional coordinate space. The ground state solution is found
by iterative relaxation. The time evolution is done by the
time-splitting method. For details see the review \cite{ownrw}. 

To compute the spectral distributions, we employ spectral analysis
after TDLDA propagation \cite{ownrw,bigax1}.  This technique which had
been developed and much applied for studying dipole resonances
\cite{ownrw,bigax1,yabana} is equally well adapted to the present case
of the scissors mode. The dedicated initialization of the scissors
channel is achieved by rotating the ground-state electron cloud by a
small angle out of the equilibrium. The subsequent dynamics is analyzed
with the orbital angular moment $<\hat{L}_i>$ which is the appropriate
observable for the orbital M1 strength. Crosstalk between the principal
axes of the cluster is very small. Thus rotation about the $x$-axis
(using $\hat{L}_x$ as a generator) explores the dynamics of
$\bar{L}(t)=\langle\hat{L}_x\rangle$ and similarly for $y$ and $z$. We
follow the time evolution for 200 fs. The Fourier transform of
$\bar{L}(t)$ provides the spectral strength distribution of orbital M1
strength. We also study the crosstalk to other modes. This is done
simply by taking a protocol of the dipole moments $\hat{D}_i\propto
r_i$ as well as spin-dipoles $S_i\propto r_i\hat{\sigma}_0$ (where
$\hat{\sigma}_0$ is the Pauli spin matrix along quantization
direction) and Fourier transforming them.  In that case, we consider
the spectral power distribution which is the sum of squared absolute
values of the respective Fourier transforms. For a more detailed
discussion of the spin-dipole see Ref. \cite{spin_na12}.

The global deformation of the cluster ionic background is crucial for
the scissors excitations \cite{lipparini_PRL,Ne_PRL}. It is 
characterized by the dimensionless quadrupole moments recoupled to 
the total deformation $\beta$ and triaxiality $\gamma$:
\begin{eqnarray}\label{eq:deform}
 && \beta = \sqrt{\beta_{20}^2+2\beta_{22}^2}
    \quad,\quad
    \gamma = \mbox{atan}\frac{\sqrt{2}\beta_{22}}{\beta_{20}}
    \quad,\quad
\nonumber
\\
 &&
 \beta_{2m}
 =
 \frac{4\pi}{5N_I} \frac{\overline{r^2Y_{2m}}}{\overline{r^2}}
\end{eqnarray}
where $N_I$ is the number of ions (atoms). The definition of
the $\beta_{2m}$ contains a simple classical averaging over the ionic
positions.

In order to check the impact of ionic structure, we also use, as an
alternative to the detailed ionic background, the soft jellium model
\cite{SJ} for the ionic density
\begin{eqnarray} \label{eq:densi}
 \rho_I({\bf r})
 &=&
 \frac{\rho_{I0}}{1+exp((r-R(\Theta ))/\alpha )}
 \quad,
\\
\label{eq:R}
  R(\Theta )
  &=&
  R_0\left(1+\sum_{m=0,\pm2} \delta_{2m}Y_{2m} (\Theta )\right)
\end{eqnarray}
where $R_0=Cr_s N_I^{1/3}$, $r_s=3.96\,{\rm a}_0$ is the
Wigner-Seitz radius, the coefficient $C$ is adjusted to ensure volume
conservation $\int d{\bf r} \rho_i({\bf r})=N_I$, 
$\rho_{I0}=3/(4\pi r_s)$ is the bulk density.  The diffuseness
$\alpha$ of the jellium surface allows to achieve a good reproduction
of the empirical optical properties \cite{SJ,alonso_diff}. It can be
justified by folding a steep jellium drop with a local ionic
pseudo-potential \cite{SJ}.  We choose the diffuseness $\alpha
=1\,{\rm a}_0$. The generating deformation parameters $\delta_{20}$
and $\delta_{22}=\delta_{2-2}$ are adjusted to give the intended
deformations $\beta$ and $\gamma$ according to Eq.
(\ref{eq:deform}). The averaging in Eq. (\ref{eq:deform}) then becomes a
weighted integration over the jellium density $\rho_I$.

\section{Basic aspects of scissors mode}

Before proceeding to the fully detailed TDLDA calculations, it is
worthwhile to look at macroscopic features of the scissors mode
as well as to present its simple microscopic view in 
terms of a deformed shell model.

\subsection{Collective picture of rotational vibrations}

In a collective picture, the scissors mode can be viewed as
small-angle rotations of the spheroid of valence electrons against the
spheroid of ions (see Fig. 1 
, part a)).
More precisely, the displacement field of this mode consists out of a
rigid rotation of the electrons against the positive ionic background
plus a compensating quadrupole term which serves to tune vanishing
velocity at the surface \cite{lipparini_PRL}
(see Fig. 1 
, part b)).:
\begin{equation}
{\vec u}({\vec r})={\vec \Omega}\times {\vec r}+ \beta
(1+\beta/3)^{-1}%
{\bf \nabla}(yz)  
\quad .
\label{zisp}
\end{equation}
The distortion of the momentum Fermi sphere generates a restoring
force leading to rotational oscillations which constitute the scissors
mode \cite{lipparini_PRL}.

The mode is characterized by low-energy oscillations with strong
magnetic dipole transitions to the ground state. In axial clusters,
the mode is specified by the states $|\Lambda^{\pi }=1^{+}>$ where
$\Lambda$ is the projection of the orbital moment onto the symmetry
axis z and $\pi $ is the space parity. The corresponding energy and
magnetic strength are \cite{lipparini_PRL,Ne_PRL}
\begin{equation}
\omega _{M1} \simeq \frac{2}{r_{s}^{2}}N_{e}^{-1/3}\beta 
\label{eq:om}
\end{equation}
\begin{eqnarray}
B(M1) &=&4\langle 1^{+}\mid \hat{L}_{x} \mid 0\rangle^2\mu _{b}^{2} 
\nonumber\\
 &=&\frac{2}{3}\omega_{M1}N_e \overline{r^2} \ \mu _{b}^{2}
\\
 &\simeq& N_{e}^{4/3}\beta \ \mu _{b}^{2}
\nonumber
\label{eq:B(M1)}
\end{eqnarray}
where $N_e$ is the number of valence electrons (we use here
natural units $m_e=\hbar =c=1$). 
The value $B(M1)$ stands for summed strength of the degenerated x- and
y-branches. The z-branch has vanishing strength for symmetry
reasons. It is worth noting that $B(M1)$ does not depend on $r_s$ and
so is the same for different metals.

In triaxial clusters, the scissors mode splits into three
branches with the frequencies \cite{lipparini_PRL}
\begin{equation}
\label{eq:omtriax}
\omega^i_{M1}=\omega_{M1}
\pmatrix{
\cos\gamma +\frac{1}{\sqrt{3}}\sin\gamma  \cr
\cos\gamma -\frac{1}{\sqrt{3}}\sin\gamma \cr
\frac{2}{\sqrt{3}}sin\gamma } 
\end{equation}
and the strengths
\begin{eqnarray}
  B^i(M1)
  &=&
  \frac{1}{3}\omega^i_{M1}N_e \overline{r^2} \mu _{b}^{2}
\nonumber\\
  &=&
  \frac{1}{2}B(M1)
  \pmatrix{
   \cos\gamma +\frac{1}{\sqrt{3}}\sin\gamma  \cr
   \cos\gamma -\frac{1}{\sqrt{3}}\sin\gamma   \cr
  \frac{2}{\sqrt{3}}sin\gamma  } 
\label{eq:Btriax}
\end{eqnarray}
where vectors run over $i\!=\!x,y,z$ components. 
The estimates show that both the frequencies and strengths are
proportional to deformation $\beta$, which reflects the fact that
collective scissors modes can exist only in deformed systems. This
feature may serve as an additional indicator of quadrupole deformation
in clusters. The scissors mode has been already observed in several
systems with different physical nature (nuclei, Bose condensate,
quantum dots). This means that the mode is a general dynamical
phenomenon persistent for deformed two-component systems. It is
interesting to note that there is another kind of universal orbital
magnetic mode, the twist M2 mode, which has been recently predicted in
clusters \cite{twistPRL}.  The twist may exist in systems of any shape
and it is the strongest multipole magnetic mode in medium and large
{\it spherical} alkali metal clusters \cite{twistPRL}.

\subsection{M1 transitions in the deformed shell model}
\label{sec:shelmod}

In order to see which single electron states are the most relevant in
building the scissors mode, we check the structure of the transition
in the framework of the Clemenger-Nilsson model
\cite{NilCle,DeHeer}.
The major shells ${\cal N}$ of the spherical oscillator remain a
useful sorting scheme for slightly deformed clusters.  
Taking into account that ${\cal N}$ 
is connected with the space parity of the levels
inside the shell as $\pi =(-1)^{\cal N}$,  
one may immediately conclude that  
$\Lambda^{\pi}=1^+$ scissors mode has to be mainly generated by low-energy
$\Delta {\cal N}=0$ transitions inside the valence shell and  
high-energy $\Delta {\cal N}=2$ transitions through two shells.   

More information can be obtained from the expression
for the orbital M1 transition element which
in axially deformed clusters has the form
 \begin{eqnarray}
  && \langle\Psi_{p}|{\hat L}_{+1}|\Psi_{h}\rangle= 
 {\textstyle\frac{1}{2}}
 \delta^{\mbox{}}_{\pi_{p},\pi_{h}}\delta^{\mbox{}}_{\Lambda_{p},\Lambda_{h}\!+\!1}
 \label{eq:me}\\
  &&
 \qquad\qquad
 \sum_{nL} a^{p}_{nL}a^{h}_{nL}\sqrt{L(L\!+\!1)\!-\!\Lambda_h(\Lambda_h\!+\!1)} \quad .
 \nonumber
 \end{eqnarray}
 Here, the wave function of a particle ($\nu =p$) or hole ($\nu =h$)
 deformed state $\Psi_{\nu ={\cal N}n_z\Lambda}=\sum_{nL}
 a^{\nu}_{nL}R_{nL}(r) Y_{L\Lambda}(\Omega)\chi_{1/2\nu}$ is written as
 a superposition of spherical $(nL)$-configurations in which $n$ stands
 for the number of radial nodes; $n_z$ labels the projection 
 of the principle shell number ${\cal N}=n_x+n_y+n_z$ into $z$-axis.   
 Eq. (\ref{eq:me}) shows that  the scissors mode is generated by 
$\Lambda_p=\Lambda_h\pm 1$ transitions
between wave function components belonging to one and the same
spherical $(nL)$-configuration. In spherical systems 
$(nL\Lambda)$-states belonging to such configuration are degenerate
while in deformed systems these states are energetically split
and so may be connected by M1 transitions 
with non-zero energies. This is just the origin of the scissors mode.  
Obviously, the energy scale of the scissors mode is determined 
by the deformation energy splitting and so is rather
small. This explains predominantly low-energy ($\Delta \cal{N}$=0)
character of the scissors mode. Just the low-energy branch exhausts 
most of the scissors $B(M1)$ strength.   
The high-energy ($\Delta \cal{N}$=2)
branch of the mode carries much weaker strength since the 
particle states involved into ($\Delta \cal{N}$=2) transitions 
include only small fractions of $(nL)$-configurations from the valence 
shell. 

One may show
\cite{Ne_PRL} that in the harmonic oscillator space the 
M1 transition matrix element
is proportional to the matrix element of the quadrupole
operator $r^2Y_{21}$.  This indicates a coupling between the high-energy 
($\Delta \cal{N}$=2) branch of scissors mode and electric 
$\lambda\mu=21$ quadrupole plasmon in deformed clusters \cite{Ne_PRL}.
  
Note that in spherical clusters the M1 transitions connect
degenerated states from one and the same $(nL)$-configuration.
These transitions have zero energy and, therefore, the scissors mode  
in spherical {\it jellium} clusters is absent.  The ionic structure 
destroys the spherical symmetry even at
zero global deformation $\beta=0$ and thus allows some ${\hat L}_{\pm
1}$ -response in such clusters. It remains to be seen how strong this
effect is in actual calculations.

Eq. (\ref{eq:me}) hints that the strength of the scissors modes 
increases with the orbital moments $L$ which are involved.
Besides that, low azimuthal quantum numbers 
$\Lambda$ favor the mode.  In heavy clusters high orbital moments 
are accessible and thus these clusters can 
exhibit strong orbital effects \cite{Ne_PRL}.  Already at $N_e\sim
300$ the magnetic strength $B(M1)$ can reach impressive value of
300-400 $\mu _{b}^{2}$.  The RPA calculations confirm the trend
(\ref{eq:om}) for the frequency but reveal in light clusters
considerable fluctuations around the trend (\ref{eq:B(M1)}) for the
strength \cite{Ne_PRL,Ne_next}. These fluctuations are due to quantal
shell effects.  Finally, it is worth to mention that the scissors mode
determines van Vleck paramagnetism and leads to strong anisotropy of
orbital magnetic susceptibility 
\cite{lipparini_PRL,Ne_PRL,Ne_next,Ne_Rich}.

\subsection{Preparatory example}

As a first example, we consider the simple case of a strictly axial
cluster. To this end we chose ${{\rm Na}_{11}}^+$ with a soft jellium
background (generating deformation $\delta_{20}=0.38, \gamma =0$).
The symmetry allows only degenerate $x$- and $y$-scissors modes.  The
associated spherical valence shell (2s,1d) consists of 4 levels:
${\cal N}n_z\Lambda =$ 220, 211, 202 and 200.  In ${{\rm Na}_{11}}^+$,
the Fermi level 220 is fully occupied and the other three levels
remain empty.  The only possible scissors transition is $220
\rightarrow 211$.  The low density of particle-hole ($1ph$) excitations
reduces fragmentation of the M1 strength 
and thus one has only one pronounced scissors peak, the ideal 
scenario for a first exploration.
The upper panel of Fig. 2 
shows the spectral
distribution of M1 strength. The full line is the result from the
linearized TDLDA calculations \cite{Ne_next,Ne_AP} with full
interaction. We use this clean test case to identify the various contributions
of the residual interaction. The ground state is always computed with
the full energy functional. But for the excitation spectrum, we
consider also the terms of the residual interaction separately. One
finds in the upper panel the spectrum for the cases of no residual
interaction at all (unperturbed $1ph$ response) and of pure
Coulomb. The Coulomb interaction is strongly repulsive (like for the
dipole mode) and results in a strong blue shift of the unperturbed
strength. The blue shift for the total residual interaction is much
smaller.  This is due to the strongly attractive residual interaction
from the exchange-correlation part of the functional. In fact, the
exchange-correlation interaction alone produces so much attraction
that the scissors mode acquires an imaginary frequency. The mode
becomes unstable and is thus missing in the figure.

The lower panel of Fig. 2 
shows as a complementing
information the radial profile of the residual interaction.  The
density dependent operator $Q_{21}({\vec r})$ of multipolarity
$\lambda\mu =21$ is in fact the (radially averaged) response potential
$-(\partial^2 E/\partial\rho^2) \delta\rho$ where $\delta\rho$ is the
transition density and $E$ is the energy functional. The plot shows
the full residual interaction as well as its two separate
contributions. Repulsive parts are negative in this
representation. One sees nicely that Coulomb is strongly repulsive,
exchange-correlation is attractive and the total effect is a moderate
repulsion. After all we see that the net residual interaction is
composed from two counteracting pieces. This inhibits any general
statement about the expected shift. Blue shift as well as red shift
can emerge depending on the actual transition density. But one can
predict that the total residual interaction will always be rather
small (with possible exceptions in special cases).

\section{Results and discussion}

\subsection{${\rm Na}_{12}$:   
global deformation versus ionic structure}

Figure 3 
shows for ${\rm Na}_{12}$ the scissors
strength, i.e. the Fourier transformed signal $\bar{L}_i(t)$, 
as computed from real time TDLDA (section \ref{sec:frame}).  The
spin-saturated triaxial ground state and spin-polarized axial isomer
are considered with both ionic and jellium backgrounds. Triaxiality
mixes the states with different $\Lambda$ but still conserves the
space parity. The ionic background makes the states parity mixed as
well.


The spectra for the triaxial ground state (left lower panel) show
the modes $L_x$, $L_y$, and $L_z$.  The
$z$-mode has one dominant peak at the low-energy end. Two other
modes are strongly fragmented. A quick glance at the lower insert
explains this. The density of $1ph$ states is very large in this
neutral and triaxial cluster and thus spectral fragmentation is very
likely. The lower end of the spectrum coincides with the lowest $1ph$
excitation which hints that the net effect of the residual interaction
is rather weak. A simplified axial treatment can roughly explain the
origin of three main peaks in x- and y-responses: in ${{\rm Na}_{12}}$
the Fermi level is one-half occupied level ${\cal N}n_z\Lambda =$211 
and within the valence (2s-1d)- shell
there should be three main scissors transitions: $220 \rightarrow
211$, $211 \rightarrow 202$, and $211 \rightarrow 200$.

Note also the M1 strength around 3 eV.  This is the high-energy branch
of the scissors mode. It emerges from the scissors 
transitions through two quantum shells $(\Delta{\cal N}=2)$
as mentioned in subsection \ref{sec:shelmod}. This branch 
exhibits a coupling to the $\lambda\mu =21$
component of the quadrupole plasmon \cite{Ne_PRL}.
The energy range coincides perfectly with the
position of the quadrupole plasmon around 3 eV (it is placed
approximately at $\sqrt{6/5}$ times the frequency of the Mie plasmon
which is here around 2.8 eV). The strength is strongly fragmented
because the weakly bound triaxial ${{\rm Na}_{12}}$ has  a high density of 
$1ph$ states in this energy region (actually not shown in the figure).

The upper left panel of Fig. 3
shows results for a
comparable jellium background. For this aim, we have computed the
global deformation of the ionic configuration. This yields
$\beta=0.55$ and $\gamma=17^o$. The jellium background is tuned to
have precisely the same deformation. The $1ph$ states start now at
much lower energy since the energy gap at the Fermi surface is lower
than with an explicit ionic configuration. This has two reasons: the jellium
model is less bound here and we are not perfectly in the minimum
of the jellium deformation energy surface, which does not coincide
with the deformation associated to the case with explicit ions. 
Consequently, the whole
spectra are shifted to lower frequencies where the first $1ph$ states
are now situated. This corroborates again the fact that the residual
interaction has a small effect. Moreover, we see by comparison
with the $1ph$ spectra that the residual interaction can act both
ways, repulsive or attractive. It is interesting to note that, in
spite of the strong general downshift as compared to the case with detailed
ionic background, the relation amongst the scissors states (ordering,
fragmentation) remains almost unchanged. And the downshift is simply
given by the spectral gap in the ground state. This hints that global
deformation is the crucial ingredient determining the pattern of the
scissors spectra.

The comparison of the $1ph$ spectra in the 
left column of Fig. 3 
is puzzling. It looks as if the
better bound ionic configuration has the higher density of states. 
Part of this is indeed the effect that the larger spectral gap
compresses the states just above the gap. Another part is generated by
the fact that the jellium background is reflection symmetric about all
three major planes while the ionic background induces slight symmetry
breaking. There are more degeneracy in the jellium spectrum and
none at all for the ionic background.  

The right panels of Fig. 3
show the same analysis
for the first isomer. This configuration is axially symmetric,
i.e. $\gamma=0^o$, but spin polarized with net spin 2
\cite{spin_na12}. Total deformation is $\beta=0.58$, close to the
deformation of the triaxial state. The upper right panel 
(axial jellium model) shows nicely the exact degeneracy of $x$ with $y$
modes and the absence of the $z$ mode. The spectrum is again
fragmented in accordance with the high density of $1ph$ states. The
lower right panel shows results with ionic structure. The $z$ mode
still nearly disappears. The degeneracy of $x$ and $y$ modes is somewhat
broken because a detailed ionic structure breaks strict axial
symmetry. This cluster is axial only in average, i.e. in the sense
that $\gamma=0^o$. The $1ph$ spectra are about at the same position in
both cases here and similarly the fully coupled spectra. The responses
are compatible with the assumption of small residual interaction.
Moreover, we see that the detailed ionic structure does not help the
$z$ mode to show up (which would, in principle, be possible). Again, we
conclude that global deformation determines the overall pattern of
scissors modes.

The inserts between upper and lower panels in Fig. 3
show the 
collective estimates (\ref{eq:omtriax}) and (\ref{eq:om})
for the scissors frequencies, obtained
in a jellium model with given deformation 
parameters. Although the estimates belong to 
the jellium case, they
can be connected with lower panels as well, as a collective 
view where only the global deformation counts. The estimates provide 
rough positions of the scissors modes but they miss any details. The
actual spectra show that the details, driven by interference with
$1ph$ structure, are crucial in these very soft clusters with
a high level density.

\subsection{ ${\rm Na}_{16}$: crosstalk}

The orbital M1 mode and spin excitations belong to the same
channel (see more on spin modes in clusters in Refs. 
\cite{S_spin_RPA}-\cite{SL_spin_SR}).
 The orbital strength $\sim B(M1)\propto N_e^{4/3}$ and power
spectra $\sim B(M1)^2\propto N_e^{8/3}$ evidently overrules the spin 
strength ($\propto N_e^0$) for large clusters.
But they can compete in the small clusters considered here. There is 
an intriguing question how strong is a crosstalk between the modes. 
This point is better discussed in terms of power spectra,
although we are in the linear regime. Mind that we are now plotting in
logarithmic scale which is more appropriate for power spectra with
their larger variations.

The various power spectra  emerging from excitation
of the scissors mode (i.e. with a rotational distortion)
are shown for Na$_{16}$ in the upper block  
of Fig. 4 
The block exhibits angular,  
spin-dipole and dipole responses. Both unpolarized 
ground state and spin-polarized first isomer of the cluster are considered.
This case is different from ${\rm Na}_{12}$ in that both configurations 
are triaxial. The spin-saturated ground state and spin-polarized isomer
have deformations $\beta=0.35$ with $\gamma=38^o$ and 
$\beta=0.32$ with $\gamma=15^o$, respectively.

The uppermost panels show the angular momentum spectra
associated with M1 strength.
The patterns are similar to the
previous case of ${\rm Na}_{12}$ but differ in detail due to different
$1ph$ distributions. The lowest panels of the upper blocks
show the power spectrum in the
dipole channel driven by the crosstalk with the angular momentum mode.
The coupling is obviously very weak. In fact, there would be no
crosstalk at all in a reflection symmetric jellium model. The effect
here is triggered by the slight asymmetries in the detailed ionic
configuration. 

The middle panels show the crosstalk with the spin-dipole mode.  There
is no crosstalk at all in the case of the spin saturated ground state.
But a strong crosstalk appears for the spin-polarized isomer. In that
case, the scissors and spin motions are fully coupled 
and have to be considered as one hybrid mode.  As complementing
information, we show in the lower block of Fig. 4
the spin-dipole strength after spin-dipole excitation. For the isomer
(right column), one recognizes the same spin-dipole spectrum as in
middle panel of the upper block. The strength distribution is, of
course, different because the excitation channel was different.  For
the spin saturated ground state (left column), there is now a clear
spin-dipole signal. This mode is clearly separated from the scissors,
but it resides precisely in the same low-energy range as the scissors
mode.  An experimental discrimination would require to separate the
spin and orbital currents. This question has been extensively
discussed in nuclear physics where the form factor for inelastic
electron scattering may give access to the separate information (see,
e.g., Ref.  \cite{richter}).  The problem is much worse
for clusters due to very low energies involved. The easiest way is to
exploit the scaling with system size and to go for large clusters
where spin-dipole excitations become relatively unimportant (see
discussion in Ref. \cite{Ne_next}).

It is worth to emphasize that the cross-talks discussed above connect 
the modes of the opposite space parity. This becomes possible
due to mixing configurations of a different parity in the detailed ionic 
background.

\subsection{ ${{\rm Na}_{9}}^+$: pure ionic effects}

As we have seen in the examples above, the scissors mode exists if the
electron cloud can accomplish rotational-like oscillations against a
massively deformed background. The mode does not work at all in the case
of the perfect spherical symmetry. But clusters do have ionic structure which
breaks spherical symmetry even if the global quadrupole deformation
vanishes. It is then conceivable that the electron cloud performs
rotational-like  vibrations against the ionic background. Such 
transversal optical modes exist in the solid state \cite{acustic}.  
Their frequency is proportional to momentum as in the acoustic
branch. One may expect  similar low-energy modes in a finite system
with zero global deformation as well.

To check this, we consider the example of ${{\rm Na}_{9}}^+$. Its
configuration consists in two rings of four ions toped by one single
ion.  The configuration has $\beta=0.08$ and is thus fairly close to
global spherical symmetry. It is to be noted that the top ion breaks
reflection symmetry.
Figure 5 
shows the angular momentum strength in this cluster.
There is some strength at 1.5-2 eV and another branch at 3.5-4.5 eV.  
The latter has position and strength much similar to the
high-frequency branches in the previous examples.  (The slight 
blue-shift is caused by positive net charge as compared to the neutral
clusters considered before.) The branch is again related to the 
quadrupole plasmon and 
is predominantly composed  of $\Delta{\cal N}$= 2 transitions.  However,
the low-frequency branch has higher frequency and much weaker strength
than in the previous examples.   
In fact it is related
to $\Delta\cal{N}$= 1 transitions. The larger frequency is explained
by the fact that we deal here with a small magic cluster with a big
HOMO-LUMO gap. The much lower strength is no surprise.  Note that
$\Delta\cal{N}$= 1 transitions change space parity and would not be 
accessed by M1 motion in a jellium model. But the detailed ionic background
mixes configurations of different parity and so even-parity
$\hat{L}_i$ operators can generate, though weak, the transitions of
$\Delta\cal{N}$=1 kind. This explains why, unlike the previous
examples, the first branch is weaker than the second one.

Such motion may be related to the transversal optical mode. But it is not
possible to track that down unambiguously because in this small cluster
the low-frequency M1 mode is predominantly of $1ph$ nature. Larger
magic clusters are required for such an analysis.  Altogether, we have
seen magnetic M1 strength in a ``spherical'' cluster.  This
strength has different sources than the scissors mode.  It
takes place even at zero global deformation and has very little
strength as compared to the genuine scissors modes.

\section{Conclusions}

We have presented fully microscopic TDLDA calculations for the
scissors mode in small Na clusters taking into account detailed ionic
structure, triaxiality and spin-polarization.  The results were
analyzed in terms of a simple collective (=macroscopic) picture and of
more microscopic shell-model view of the mode.  To single out the
effects of ionic background, a comparison to jellium cases was
done. Clusters with different global deformation (spherical, axial and
triaxial) were considered.
  
The test case of ${{\rm Na}_{11}}^+$ with  jellium background
displays  straightforward and unfragmented scissors peak. This case
was used to study the various contributions to the residual
interaction defining the actual peak position.  It revealed that
Coulomb and exchange-correlation terms act in opposite directions, 
resulting finally only in a small collective energy shift.  
The actual sign depends on
details, as e.g. interplay with particle-hole structures.  
The calculations for ${\rm Na}_{12}$ show that  
jellium results demonstrate good agreement with macroscopic estimates
for the low-energy scissors branch, though the estimates miss such
important aspect as fragmentation of the strength due to particle-hole
states.  There is also a high-energy scissors branch which is 
explained as a result of the transitions through two quantum shells.  
 
Triaxiality produces a richer picture.  It gives rise to an additional
$L_z$-scissors branch and to considerable increase of the
fragmentation.  Besides that, triaxiality combined with the delicate
balance between Coulomb and exchange-correlation contributions can
lead to both red and blue shifts in one and the same spectrum.

Detailed ionic structure leads to the blue-shift of the scissors mode as
compared to jellium results.  It also considerably enhances the 
fragmentation. The ionic background causes small parity-mixing which
results in a crosstalk, although weak, between positive-parity
scissors and negative-parity dipole excitations. This crosstalk takes
place for both spin-saturated and spin-polarized states.
Spin-polarized states demonstrate in addition a strong crosstalk of
the scissors and spin-dipole modes, the effect which is absent  
in spin-saturated states. Both modes happen to reside in the
same corner of low-energy states which makes their experimental
discrimination extremely tough.
   
Comparison of results from jellium with those using detailed ionic
background shows that the scissors modes are mainly determined by the
global deformation. The ionic background as such destroys locally
the spherical symmetry  and gives rise to some M1 orbital
strength (as the transversal optical modes in
solids). This becomes apparent by the existence of $L_i$ response even
in clusters with {\it zero} global deformation, like $Na_9^+$. The
strength of such a mode is an order of magnitude smaller than
the collective scissors strength which lives up in truly deformed
systems. 
 
The scissors mode is a general dynamical effect peculiar to
different Fermi (and even Bose) systems with a disturbed symmetry. 
In the particular 
case of atomic clusters, the mode demonstrates very interesting
features deserving further study. The main problem, not
tackled in this paper, is its unique experimental identification. Such
questions will be considered in a forthcoming publication.

\begin{acknowledgments} 
{The work was supported by grants from RFBR (00-02-17194),
Heisenberg-Landau program (BMBF-BLTP JINR), DFG (436RUS17/102/01)
the French-German exchange program PROCOPE (99074),
and Institut Universitaire de France. V.O.N.  
thanks N. Lo Iudice for useful discussions.}
\end{acknowledgments}

\newpage

{\large\bf Figure captions}:

\vspace{0.5cm}\indent
{\bf Figure 1}:
Macroscopic view of scisors mode \cite{lipparini_PRL}: a) rigid rotation,
b) rotation within a rigid surface.

\vspace{0.5cm}\indent
{\bf Figure 2}:
Scissors photo-absorption cross section (upper panel) and radial
profile of residual interaction (lower panel) for ${{\rm
Na}_{11}}^+$. Four cases are compared: no residual interaction at all
(denoted ph, dashed line in upper panel), only Coulomb residual
interaction (dotted lines), only exchange-correlation
 part of residual interaction
(dashed line in lower panel), full residual interaction (full line).
No scissors strength could be shown for the exchange-correlation 
part alone because in this case
the residual interaction is so strongly attractive that an
imaginary frequency emerges.

\vspace{0.5cm}\indent
{\bf Figure 3}:
Strength distribution of angular momentum modes in ${\rm Na}_{12}$.
The quadrupole deformation of the jellium background is chosen to be
the same as in the states with full ionic structure. The 
modes $L_x$, $L_y$, and $L_z$ are distinguished by the lines as
indicated in the left upper panel. 
 Inserts under the panels show the 1ph spectra 
(bars) up to $\sim$ 2 eV. The boxes in the left and right inserts 
between upper and lower panels 
show the collective estimates (\ref{eq:omtriax}) and 
(\ref{eq:om}), respectively, for the scissors 
frequencies in the jellium model. 

\vspace{0.5cm}\indent
{\bf Figure 4}:
The upper block shows
various power distributions in ${\rm Na}_{16}$ after angular
excitation: angular (upper), spin-dipole (middle), 
dipole (lower).  Unpolarized triaxial ground state (left)
and spin-polarized axial isomer (right) are considered.  
The lowest block shows spin-dipole power following spin-dipole excitation.
In all the calculations, full ionic structure is taken into account. 
Every case has three
modes in each spatial direction. They are distinguished by lines 
as indicated. 

\vspace{0.5cm}\indent
{\bf Figure 5}:
Strength distribution of angular momentum modes in ${{\rm Na}_{9}}^+$,
described with detailed ionic structure. The $1ph$ spectra at low
energies are indicated in the insert below.

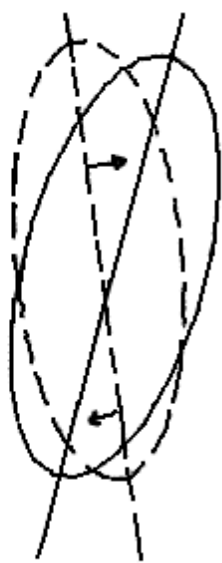
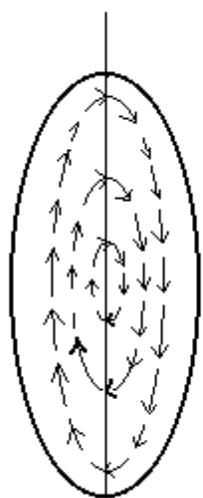
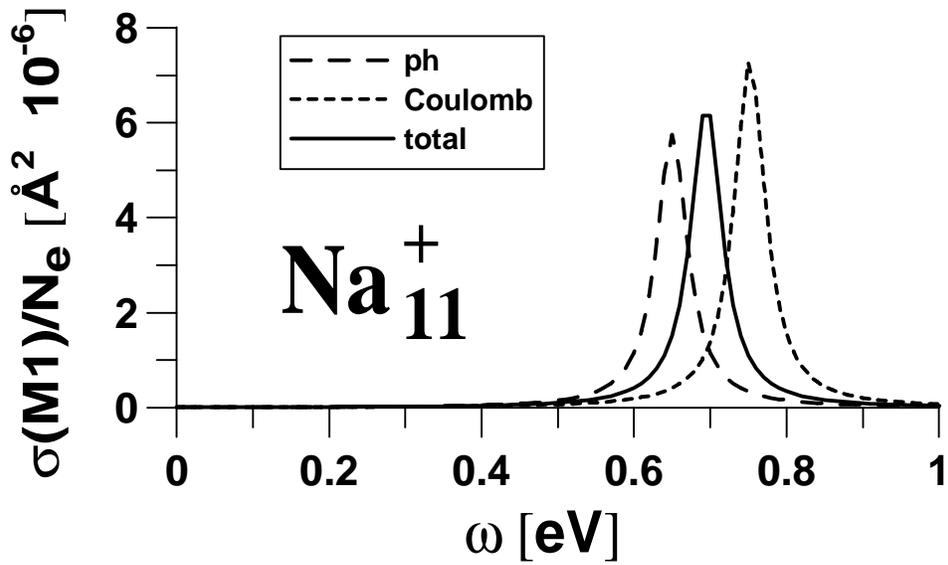
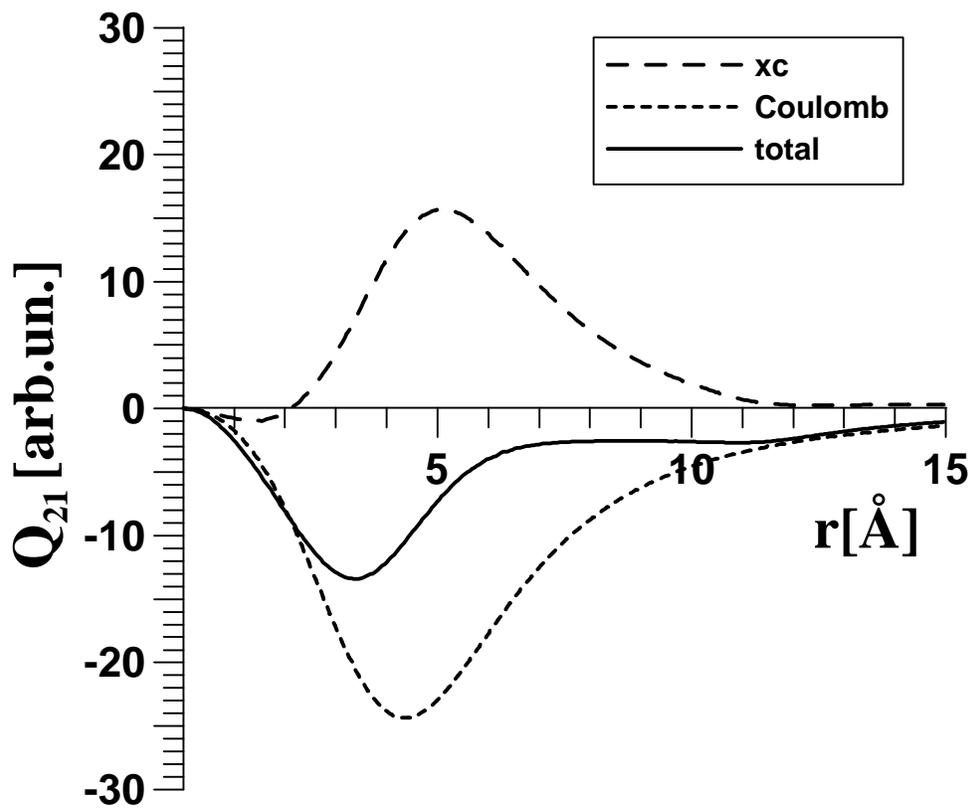
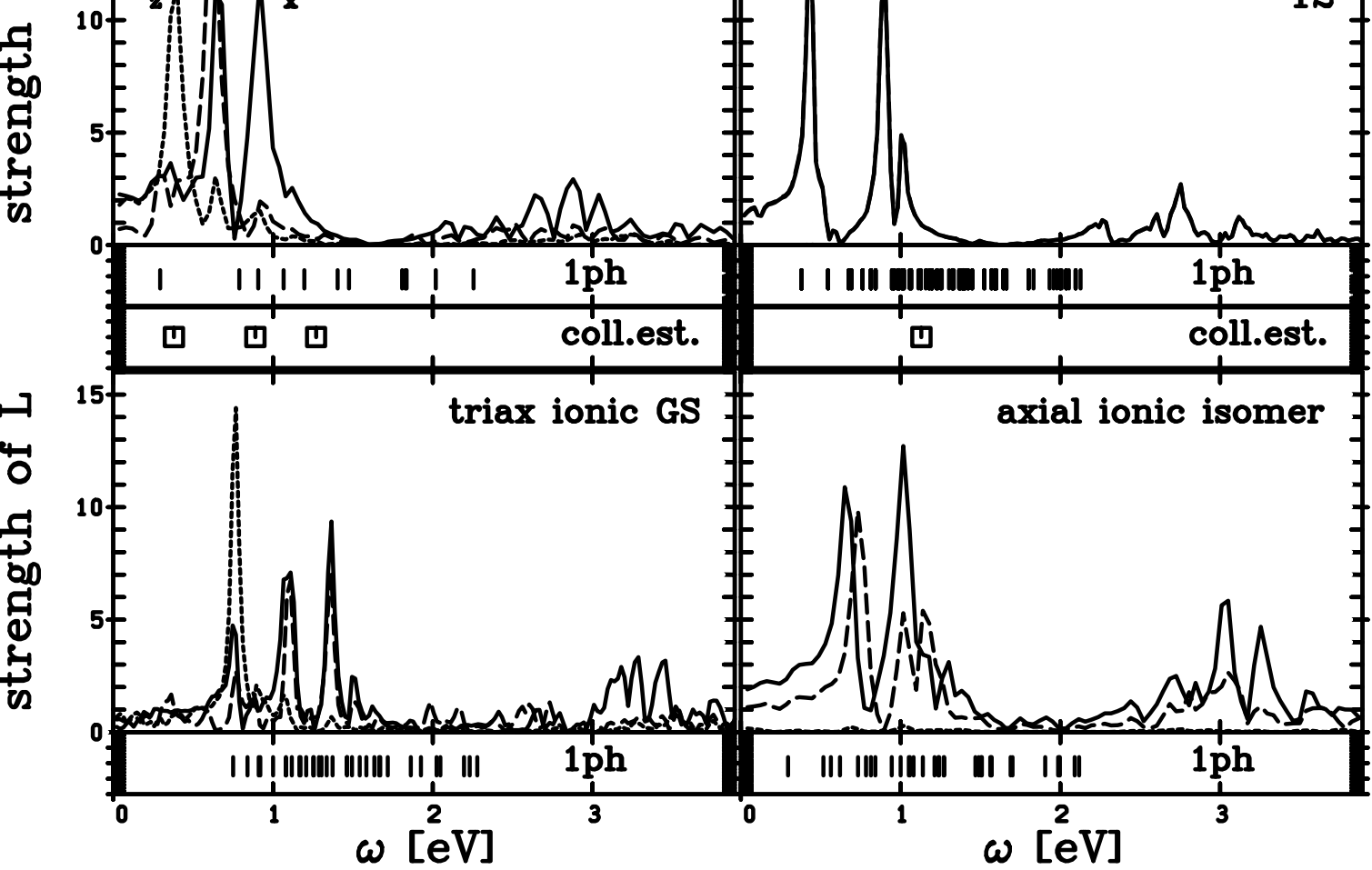
\end{document}